\documentclass[apj]{emulateapj}

\newcommand{\ii}{$i'$}

\newcommand{\zz}{$z'$}

\newcommand{\uu}{$U$}
\newcommand{\bb}{$B$}
\newcommand{\vv}{$V$}

\newcommand{\cge}{$\gtrsim$}
\newcommand{\cle}{$\lesssim$}
\newcommand{\etal}{{et\thinspace al.}}

\newcommand{\Ang}{\AA\thinspace}

\newcommand{\Ho}{$H_{0}$}

\newcommand{\tabref}[1]{Table~\ref{#1}}
\newcommand{\figref}[1]{Figure~\ref{#1}}
\newcommand{\secref}[1]{\S~\ref{#1}}
\newcommand\arcspt   {{$\buildrel{\prime\prime}\over .$}}
\newcommand\Msun     {{\ $M_{\odot}$} }

\begin{document}

\title{Surface  Brightness  Profiles of  Composite  Images of  Compact \\
Galaxies  at \boldmath{$\lowercase{z}\!\simeq\!4\!-\!6$}  in the  HUDF}

\shorttitle{Composite Surface Brightness Profiles at $z\!\simeq\!4\!-\!6$}

\author{N.  P.  Hathi\altaffilmark{1},  R. A.  Jansen\altaffilmark{2,1},
R.   A.   Windhorst\altaffilmark{2,1},   S.  H.   Cohen\altaffilmark{2}, \\
W.   C.  Keel\altaffilmark{3},   M.   R.  Corbin\altaffilmark{4}   and
R. E. Ryan Jr.\altaffilmark{1}}

\altaffiltext{1}{Department  of  Physics,  Arizona  State  University,
Tempe, AZ 85287-1504, USA}

\altaffiltext{2}{School of Earth  and Space Exploration, Arizona State
University, Tempe, AZ 85287-1404, USA}

\altaffiltext{3}{Department  of Physics  and Astronomy,  University of
Alabama, Box 870324, Tuscaloosa, AL 35487, USA}

\altaffiltext{4}{U. S. Naval Observatory, 10391 W. Naval Observatory
Road, Flagstaff, AZ 86001-8521, USA}

\email{Nimish.Hathi@asu.edu}
\shortauthors{Hathi et al}


\begin{abstract}

The Hubble  Ultra Deep Field  (HUDF) contains a significant  number of
\bb-, \vv- and  \ii-band dropout objects, many of  which were recently
confirmed to  be young star-forming  galaxies at $z\!\simeq\!4\!-\!6$.
These galaxies are too  faint individually to accurately measure their
radial surface brightness profiles.   Their average light profiles are
potentially of  great interest,  since they may  contain clues  to the
time since  the onset of  significant galaxy assembly.   We separately
co-add \vv, \ii- and \zz-band  HUDF images of sets of $z\!\simeq\!4,5$
and $6$  objects, pre-selected to have nearly  identical compact sizes
and the  roundest shapes.  From these  stacked images, we  are able to
study the average(d) radial structure  of these objects at much higher
signal-to-noise ratio  than possible  for an individual  faint object.
Here we explore the reliability and usefulness of a stacking technique
of compact  objects at $z\!\simeq\!4\!-\!6$ in the  HUDF.  Our results
are:  (1) image  stacking provides  reliable and  reproducible average
surface  brightness profiles;  (2) the  shape of  the  average surface
brightness profile  shows that even  the faintest $z\!\simeq\!4\!-\!6$
objects are  \emph{resolved}; and  (3) if late-type  galaxies dominate
the  population  of  galaxies  at  $z\!\simeq\!4\!-\!6$,  as  previous
\emph{HST}  studies have  shown for  $z\!\lesssim\!4$, then  limits to
dynamical age  estimates for these galaxies from  their profile shapes
are comparable with  the SED ages obtained from  the broadband colors.
We  also present accurate  measurements of  the sky-background  in the
HUDF and its associated 1$\sigma$ uncertainties.

\end{abstract}



\section{Introduction}\label{introduction}

In  the last  decade,  ground  and space  based  observations of  high
redshift  galaxies  have  begun  to  outline  the  process  of  galaxy
assembly.   The details of  that process  at high  redshifts, however,
remain poorly constrained.  There  is increasing support for the model
of  galaxy formation,  in  which the  most  massive galaxies  assemble
earlier      than       their      less      massive      counterparts
\citep[e.g.][]{cowi96,guzm97,koda04,mcca04}.   A detailed  analysis of
the  `fossil record'  of  the current  stellar  populations in  nearby
galaxies   selected   from  the   \emph{Sloan   Digital  Sky   Survey}
\citep[SDSS;][]{york00} provides  strong evidence for  this downsizing
picture  \citep{heav04,pant07}.   The  increasing number  of  luminous
galaxies  spectroscopically   confirmed  to  be   at  $z\!\simeq\!6.5$
\citep[e.g.][]{hu02,  kodi03,  kurk04,  rhoa04,  ster05,  tani05},  or
\cle0.9 Gyr  after the Big  Bang, also supports this  general picture.
In an  alternate hierarchical scenario, arguments have  been made that
significant number  of low luminosity  dwarf galaxies were  present at
these times,  and were the main  contributor to finish  the process of
reionization   of  the  intergalactic   medium  \citep{yan04a,yan04b}.
However,  there  is  presently  little information  on  the  dynamical
structure of  these or  other galaxies at  $z\!\simeq\!6$.  It  is not
clear  whether  these  objects  represent isolated  disk  systems,  or
collapsing spheroids, mergers or other dynamically young objects.

\citet{ravi06}  used deep, multi-wavelength  images obtained  with the
\emph{Hubble Space Telescope} (\emph{HST}) Advanced Camera for Surveys
(ACS) as part  of the Great Observatories Origins  Deep Survey (GOODS)
to analyze 2-D surface brightness distributions of the brightest Lyman
Break Galaxies (LBGs)  at $2.5\!<\!z\!<\!5$.  They distinguish various
morphologies  based on the  S\'{e}rsic index  $n$, which  measures the
shape of  the azimuthally  averaged surface brightness  profile (where
$n$=1  for exponential  disks and  $n$=4  for a  de Vaucouleurs  law).
\citet{ravi06} find that 40\% of the LBGs have light profiles close to
exponential, as seen for disk  galaxies, and only $\sim$30\% have high
$n$,  as seen  in  nearby  spheroids.  They  also  find a  significant
fraction ($\sim$30\%) of galaxies with light profiles \emph{shallower}
than  exponential, which appear  to have  multiple cores  or disturbed
morphologies, suggestive  of close  pairs or on-going  galaxy mergers.
Distinction  between  these possible  morphologies  and, therefore,  a
better estimate of the formation  redshifts of the systems observed at
$z\!\simeq\!4\!-\!6$  in  particular,  is  important for  testing  the
galaxy assembly  picture, and for  the refinement of  galaxy formation
models.

One possible technique involves the radial surface brightness profiles
of  the most  massive objects  --- those  that will  likely  evolve to
become  the massive  elliptical galaxies,  which  we see  in place  at
redshifts $z\!\lesssim\!2$  \citep{driv98,vand03,vand04}.  This can be
analytically   understood  in  the   context  of   the  \citet{lynd67}
relaxation formalism and the numerical galaxy formation simulations of
\citet{vana82},  which  describe  collisionless collapse  and  violent
relaxation as the formation mechanism for elliptical galaxies.  As the
time-scale for  relaxation is shorter in  the inner than  in the outer
parts of a galaxy, convergence toward a $r^{1/n}$-profile will proceed
from  the  inside  to  progressively  larger  radii  at  later  times.
Moreover,  \citet{korm77} has  shown that  tidal perturbations  due to
neighbors can  cause the radial surface brightness  profile to deviate
from a  pure de~Vaucouleurs  profile in the  outer parts of  a galaxy.
This implies  that the radius where surface  brightness profiles start
to deviate significantly from  an $r^{1/n}$ profile \emph{might} serve
as a ``\emph{virial  clock}'' that traces the time  since the onset of
the last major  merger, accretion events or global  starburst in these
objects.

Image   stacking  methods   have  been   used  extensively   on  X-ray
\citep{nand02,bran01}  and radio  \citep{geor03,whit07} data  to study
the mean properties (e.g. flux, luminosity) of well-defined samples of
sources  that are  otherwise too  faint to  be  detected individually.
\citet{pasc96} applied  such a  stacking method to  a large  number of
optically very  faint, compact objects at $z\!=\!2.39$  to trace their
``average''   structure.    This   approach   was  also   applied   by
\citet{zibe04} to  detect the presence  of faint stellar  halos around
disk  galaxies selected  from  the  SDSS.  An  attempt  to apply  this
technique  to  high  redshift   galaxies  in  the  Hubble  Deep  Field
\citep[HDF;][]{will96}  was  not  conclusive  (H.   Ferguson;  private
communication) due to the  poorer spatial sampling and shallower depth
of   the    HDF   compared   to   the   Hubble    Ultra   Deep   Field
\citep[HUDF;][]{beck06}.

In this paper, we use  the exceptional depth and fine spatial sampling
of the HUDF  to study the potential of  this image stacking technique,
and will estimate limits to dynamical ages of faint, young galaxies at
$z\!\simeq\!4\!-\!6$.  The HUDF  reaches $\sim$1.5~mag deeper than the
equivalent  HDF  exposure in  the  \ii-band,  and  has better  spatial
sampling than the HDF.  The  HUDF depth also allows us to characterize
the sky background very accurately, which is critical for successfully
using  a  stacking  method  to  measure  the  mean  surface-brightness
profiles for these faint young galaxies.

This  paper  is  organized  as follows:  In  \secref{observations}  we
summarize the HUDF observations, and in \secref{sample} we discuss the
selection   of    our   $z\!\simeq\!4,5$   and    $6$   samples.    In
\secref{analysis}  we  describe  our  data  analysis,  which  includes
accurately  measuring the 1$\sigma$  sky-subtraction error,  the image
stacking method to generate  mean surface-brightness profiles, and our
test of  its reliability.  In \secref{results} we  present and discuss
our  results in terms  of the  average surface-brightness  profiles of
$z\!\simeq\!4\!-\!6$ galaxies, and  in \secref{conclusion} we conclude
with a summary of our results.

Throughout  this paper we  refer to  the \emph{HST}/ACS  F435W, F606W,
F775W,  and F850LP  filters as  the \bb-,  \vv-, \ii-,  and \zz-bands,
respectively.  We assume a \emph{Wilkinson Microwave Anisotropy Probe}
(WMAP)  cosmology   of  $\Omega_m$=0.24,  $\Omega_{\Lambda}$=0.76  and
\Ho=73~km~s$^{-1}$~Mpc$^{-1}$, in  accord with the  most recent 3-year
WMAP results  of \citet{sper07}.  This  implies a current age  for the
Universe  of 13.65~Gyr.   All magnitudes  are given  in the  AB system
\citep{oke83}.


\section{Observations}\label{observations}

The HUDF  contains \cge100 objects that are  \ii-band dropouts, making
them candidates for  galaxies at $z\!\simeq\!6$ \citep{bouw04, bouw06,
bunk04, yan04b}.   Similarly, there are  larger numbers of  objects in
the  HUDF  that are  \bb-band  dropouts  (415  in total)  or  \vv-band
dropouts  (265   in  total),  and  are  candidates   for  galaxies  at
$z\!\simeq\!4$  and $z\!\simeq\!5$, respectively.   \citet{beck06} and
\citet{bouw07} find  similar number of  \bb- and \vv-band  dropouts in
the  HUDF.  A significant  fraction of  these objects  to AB\cle27~mag
have  recently  been  spectroscopically  confirmed to  have  redshifts
$z\!\simeq\!4\!-\!6$ through  the detection of  Ly$\alpha$ emission or
identifying their  Lyman break  \citep{ malh05,dow07}. We  discuss our
detailed  drop-out   selection  criteria  below.   Despite  the  depth
(AB\cle29.5~mag)  of the  HUDF images,  however, these  objects appear
very  faint and with  little, if  any, discernible  structural detail.
Visual  inspection of all  these objects  shows their  morphologies to
divide into four broad  categories: symmetric, compact, elongated, and
amorphous.


\section{Sample Selection}\label{sample}

\begin{figure}
\epsscale{1.0}
\plotone{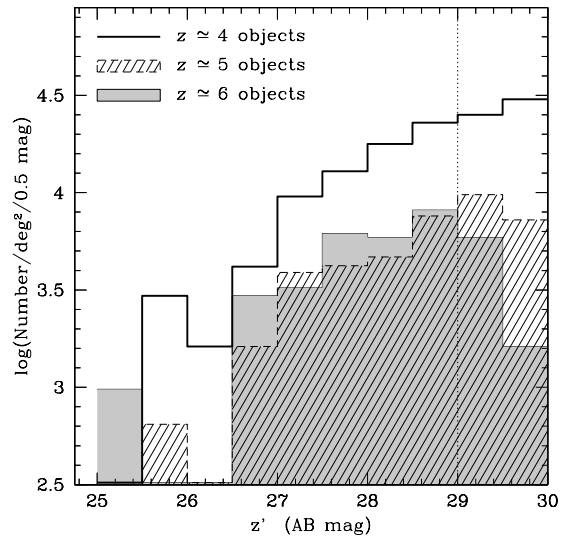}
\caption{The HUDF  number counts for all $z\!\simeq\!4,  5, 6$ objects
before  the sub-selection  of  compact isolated  $z\!\simeq\!4, 5,  6$
objects was  made.  The  vertical dotted line  shows the  magnitude to
which the number counts of  all these redshifts are complete. The area
of  the  HUDF  is  3.15$\times$10$^{-3}$  deg$^2$.}\label{fig1}
\end{figure}

We  construct  three separate  catalogs  for these  $z\!\simeq\!4,5,6$
galaxy candidates, selecting  only the \emph{isolated}, \emph{compact}
and \emph{symmetric} galaxies. We  exclude objects with obvious nearby
neighbors, to  avoid a bias  due to dynamically disturbed  objects and
complications    due   to   chance    superpositions.    \figref{fig1}
demonstrates  that  our  completeness  limit  for  $z\!\simeq\!4$  and
$z\!\simeq\!5$  objects  is  AB\cle29.3~mag,  and  for  $z\!\simeq\!6$
objects  it  is AB\cle29.0~mag.   Therefore,  all  three catalogs  are
complete  to  AB\cle29.0~mag,  which  is  equivalent  to  at  least  a
10$\sigma$ detection  for objects that  are nearly point  sources. For
each   object  in   our  $z\!\simeq\!4,5,6$   samples,   we  extracted
51$\times$51 pixel postage stamps  (which at 0\farcs03 pix$^{-1}$ span
$1\farcs53$ on a  side) from the HUDF \vv,  \ii\thinspace and \zz-band
images, respectively.  Each postage stamp was extracted  from the full
HUDF, such that the centroid of an object (usually coincident with the
brightest pixel) was at the center of that stamp.

\subsection{\boldmath {The $z\!\simeq\!4$ and $z\!\simeq\!5$ Objects (\bb-, \vv-band dropouts)}}

We used the \ii-band  selected $BVi'z'$ HUDF catalog \citep{beck06} to
select  the  $z\!\simeq\!4$  and  $z\!\simeq\!5$  objects.   With  the
\texttt{HyperZ} code \citep{bolz00},  we computed photometric redshift
estimates, using the magnitudes and associated uncertainties tabulated
in   the   HUDF    catalog.    All   objects   with   3.5$\leq\!z_{\rm
phot}\!\leq$4.5 were assigned to the bin of $z\!\simeq\!4$ candidates,
and all  objects with 4.5$\leq\!z_{\rm  phot}\!\leq$5.5 to the  bin of
$z\!\simeq\!5$ candidates.

We  then   applied  color  criteria,  similar  to   those  adopted  by
\citet{giav04},   to   select  the   \bb   ($z\!\simeq\!4$)  and   \vv
($z\!\simeq\!5$) dropout samples.  For \bb-band dropouts, we require:
\begin{displaymath}
      \left\{ \begin{array} {ll}
        (B-V) \ge 1.2 + 1.4 \times (V-z') \quad \hbox{mag} \\ 
        \hbox{and}\quad (B-V) \ge 1.2  \quad \hbox{mag} \\
        \hbox{and}\quad (V-z') \le 1.2 \quad \hbox{mag} 
		\end{array} \right.
\end{displaymath} 
For \vv-band dropouts, the following color selection was applied:
\begin{displaymath}
      \left\{ \begin{array} {ll}
	(V-i') > 1.5 + 0.9 \times (i'-z') \; \; \hbox{or} \; \; (V-i') > 2.0  \quad \hbox{mag} \\
        \hbox{and}\quad (V-i') \ge 1.2  \quad \hbox{mag} \\
	\hbox{and}\quad (i'-z') \le 1.3 \quad \hbox{mag} 
                \end{array} \right.
\end{displaymath}
We  note, that  only objects  satisfying \emph{both}  color \emph{and}
photometric   redshift  criteria   were  selected   in   our  samples.
\citet{vanz06}  using  VLT/FORS2  observed  $\sim$100 \bb-,  \vv-  and
\ii-band  dropout  objects in  the  Chandra  Deep  Field South  (CDFS)
selected based on above mentioned color criteria \citep{giav04}.  They
have spectroscopically confirmed $>$90\% of their high redshift galaxy
candidates.   Therefore, we expect  only a  small number  ($<$10\%) of
contaminants in  our sample  of dropouts.  One  or two objects  in our
final  sample  could be  such  contaminants,  but  because we  have  3
different  realizations  of  10  objects (3$\times$10),  each  showing
similar profiles, they do not appear to affect our results.

\begin{figure}
\epsscale{1.0}
\plotone{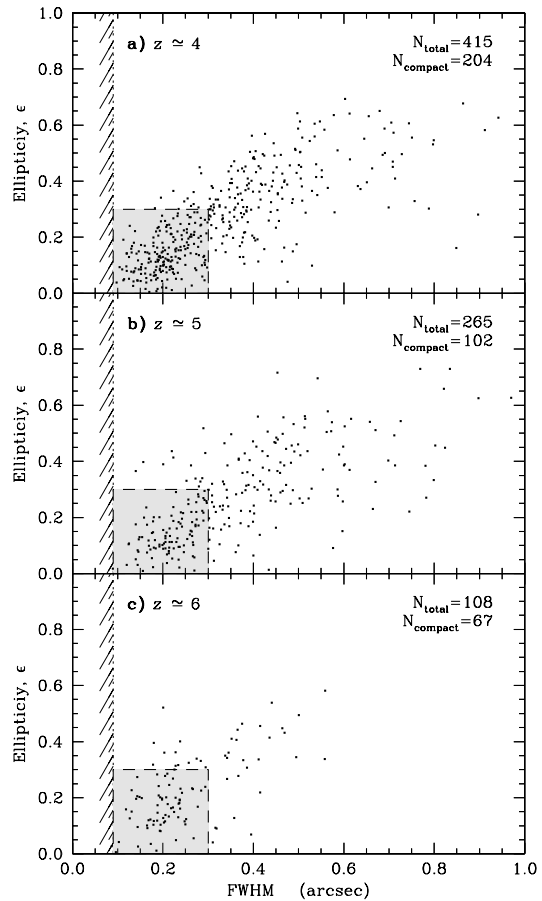}
\caption{Ellipticity,   ($1-b/a$),  versus   object   FWHM,  for   all
$z\!\simeq\!4$   {\bf   {(a)}},   $z\!\simeq\!5$   {\bf   {(b)}}   and
$z\!\simeq\!6$ {\bf {(c)}} objects selected in the HUDF.  Measurements
were  performed  in  \ii-band  for $z\!\simeq\!4$  and  $z\!\simeq\!5$
objects, while  we used the \zz-band for  $z\!\simeq\!6$ objects.  The
FWHM of a stellar image/PSF  is $\sim$3 pixels or 0\farcs09, indicated
by the leftmost hatched area  in each panel. Objects within the shaded
area meet  our additional  selection criteria on  roundness ($\epsilon 
\le   0.3$)    and   compactness   (FWHM   $\le    0\farcs3$   or   10
pixels).}\label{fig2}
\end{figure}

The $z\!\simeq\!4$  sample has  415 objects, while  the $z\!\simeq\!5$
sample has 265 objects.   In \figref{fig2}ab, we show the distribution
of the  FWHM and ellipticity, $\epsilon=(1-b/a)$, measured  in each of
the two samples  using \texttt{SExtractor} \citep{bert96}.  We further
constrained  our samples  by  imposing limits  on  compactness and  on
roundness of FWHM $\le 0\farcs3$  and $\epsilon \le 0.3$.  Again, this
is   to  minimize  the   probability  that   the  $z\!\simeq\!4\!-\!5$
candidates  are significantly dynamically  disturbed, and  to maximize
the probability of selecting  physically similar objects.  Our goal is
to find  the visibly most  symmetric, least disturbed systems  for the
current  study.    This  sub-selection  leaves  204   objects  in  the
$z\!\simeq\!4$ sample  and 102  objects in the  $z\!\simeq\!5$ sample.
Most of these  objects are faint, and are only a  few pixels across in
size, and,  hence, have larger uncertainties in  their measurements of
FWHM and ellipticity.  Therefore, we also checked our objects visually
to   eliminate  any   possibility  of   our  selected   objects  being
contaminated by  unrelated nearby objects, being  clearly extended, or
objects with complex morphologies.

\subsection{\boldmath {$z\!\simeq\!6$ Objects (\ii-band dropouts)}}

\citet{yan04b} found 108  possible 5.5$\leq\!z\!\leq$6.5 candidates in
the HUDF to $m_{AB}$($z_{850}$)=30.0~mag. \citet{bunk04} independently
found  the brightest  54  of these  108  $z\!\simeq\!6$ candidates  to
AB=28.5~mag.  Similarly, deep \emph{HST}/ACS grism spectra of the HUDF
\ii-band dropouts confirm \cge90\%  of these objects at AB\cle27.5 mag
to be  at $z\!\simeq\!6$ \citep{malh05,hath07}.  Using  the catalog of
\citet{yan04b},  we extracted  108 postage  stamps,  each 51$\times$51
pixels in size, from the HUDF \zz-band image.

Like  for  the $z\!\simeq\!4$  and  $z\!\simeq\!5$  objects, for  each
$z\!\simeq\!6$ object  we measured  its \zz-band FWHM  and ellipticity
using   \texttt{SExtractor}.    \figref{fig2}c   shows  the   measured
ellipticity  versus FWHM  for  all 108  $z\!\simeq\!6$ candidates.   A
smaller sample of 67 objects satisfies our constraints on the FWHM and
ellipticity.  Further visual inspection,  to make sure that our sample
has only isolated, compact and round objects, leaves 30 objects in our
$z\!\simeq\!6$  sample.  We  therefore  imposed a  sample  size of  30
objects also on the two lower redshift bins after visual inspection.

The  results  in  this  paper  are therefore  based  on  approximately
(30/415)$\sim$7\%,  (30/265)$\sim$11\%, and (30/108)$\sim$28\%  of the
total $z\!\simeq\!4,5$ and $6$ galaxy populations.


\section{Results}\label{analysis}

\subsection{The HUDF Sky Surface-Brightness Level and its rms Variation}\label{skyerror}

For  the  present  work,  it  is \emph{critical}  that  we  accurately
characterize  the  sky-background, and  correctly  propagate the  true
1$\sigma$ errors due to the subtraction of this sky-background. In the
following, we  will pursue  two complimentary approaches  to determine
the sky surface-brightness, and  compare the results. Here, we discuss
the \zz-band measurements in detail.

We first measured the sky-background in each of the 415 $z\!\simeq\!4$
object  stamps  (`local'  sky  measurements).   The  Interactive  Data
Language  (IDL\footnote{IDL  Website http://www.ittvis.com/index.asp})
procedure  \texttt{SKY/MMM.pro}\footnote{Part  of  the  IDL  Astronomy
User's Library,  see: http://idlastro.gsfc.nasa.gov/homepage.html} was
used to  measure the sky-background.   This procedure is  adapted from
the \texttt{DAOPHOT} \citep{stet87} routine of the same name and works
as follows.   First, the average and  sigma are obtained  from the sky
pixels.  Second,  these values are  used to eliminate outliers  with a
low  probability.   Third, the  values  are  then  recomputed and  the
process is repeated up to  20 iterations.  If there is a contamination
due to an object, then the contamination is estimated by comparing the
mean and median of the remaining sky pixels to get the true sky value.
The output of this procedure is the modal sky-level in the image.

\begin{figure}
\epsscale{1.0}
\plotone{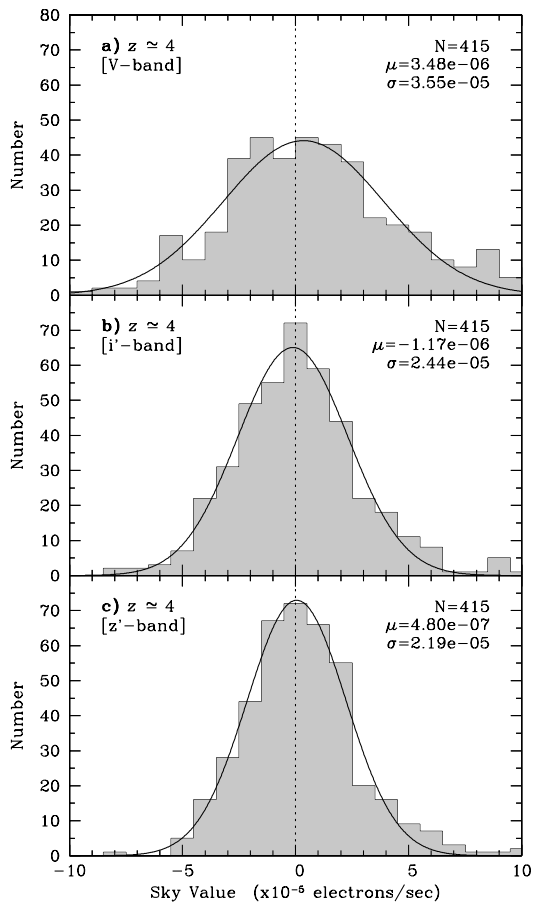}
\caption{Distribution  of  the  modal  sky background  level  used  to
estimate the 1$\sigma$  uncertainty in that level, as  measured in the
415  $z\!\simeq\!4$ \emph{object stamps}  extracted from  the drizzled
HUDF images  {\bf {(a)}}  for \vv-band, {\bf  {(b)}} for  \ii-band and
{\bf {(c)}} for  \zz-band.  The mean ($\mu$) and  the sigma ($\sigma$)
of the  best-fit Gaussian  to these distributions  are also  shown in
each panel.}\label{fig3}
\end{figure}

\begin{figure}
\epsscale{1.0}
\plotone{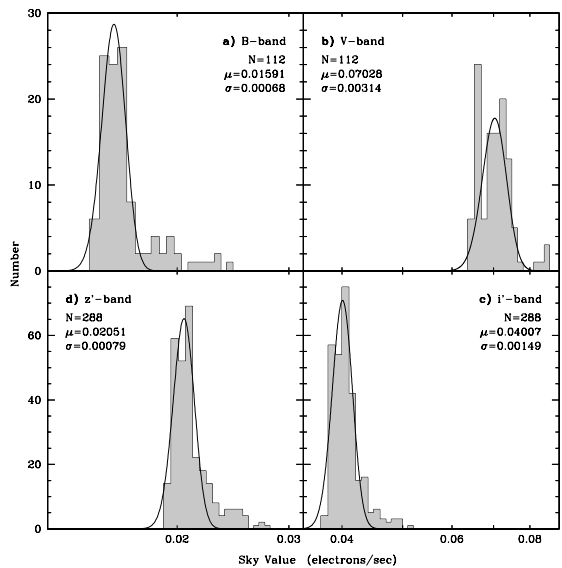}
\caption{The \emph{actual sky values} measured using header parameters
MDRIZSKY and EXPTIME from  flat-fielded HUDF exposures.  {\bf (a)} for
\bb-band  using  112 exposures.   {\bf  (b)}  for  \vv-band using  112
exposures.  {\bf (c)} for \ii-band using 288 exposures.  {\bf (d)} for
\zz-band  using  288  exposures.   The  mean  ($\mu$)  and  the  sigma
($\sigma$) of the best-fit  Gaussian to these distributions are shown
in each panel.}\label{fig4}
\end{figure}

\figref{fig3}c  shows a  histogram of  the \zz-band  modal  sky values
obtained from all  415 object stamps extracted from  the drizzled HUDF
images.   The 1$\sigma$  uncertainty in  the sky,  $\sigma_{\rm sky}$,
determined   from    a   Gaussian    fit   to   the    histogram,   is
$2.19\times10^{-5}$  electrons   sec$^{-1}$  in  the   \zz-band.   The
sky-background level  within the HUDF  was obtained from  the original
flat-fielded ACS images, because the final co-added HUDF data products
are sky-subtracted.   The header parameters MDRIZSKY  and EXPTIME were
used to  obtain the actually  observed sky-value. MDRIZSKY is  the sky
value  in   electrons  ($e^-$)  computed  by   the  MultiDrizzle  code
\citep{koek02}, while EXPTIME is the total exposure time for the image
in seconds, so that the average sky-value in the HUDF has the units of
$e^-$   sec$^{-1}$.   \figref{fig4}d  shows   the  histogram   of  the
sky-values  obtained from  288 HUDF  \zz-band  flat-fielded exposures.
The average  value of  the sky background,  I$_{\rm sky}$,  is 0.02051
$e^-$  sec$^{-1}$ pix$^{-1}$.   That  sky-value is  measured from  the
flat-fielded  individual  ACS images  with  pixel  sizes of  0\farcs05
pix$^{-1}$  and  hence, in  the  following  calculations, the  average
sky-value is  multiplied by  a factor of  (0.030/0.05)$^2$=0.60$^2$ to
obtain the corresponding average sky-value for the HUDF drizzled pixel
size of  0\farcs030 pix$^{-1}$.  Using  these values, we  estimate the
relative rms random sky-subtraction error as follows:
\[
\Sigma_{\rm ss,ran} = \frac{\sigma_{\rm sky,ran}}{{\rm I}_{\rm sky}} = \frac{2.19\times10^{-5}}{2.05\times10^{-2}\ \cdot\ 0.60^2} = 2.97\times10^{-3} 
\]


The measured average sky background level can then be expressed as the \zz-band
sky surface brightness as follows:
\begin{eqnarray*}
\mu_{z'} &=& 24.862 - 2.5 \cdot \log\left(\frac{0.0205\ \cdot\ 0.60^2}{0.030^2} \right) \\
         &=& 22.577 \pm 0.003 \ \hbox{mag arcsec}^{-2} \qquad\qquad\qquad\qquad\quad\null 
\end{eqnarray*}
where  24.862 is the  ACS/WFC \zz-band  AB zero-point,  and 0\farcs030
pixel$^{-1}$ is the drizzled pixel scale.  This is consistent with the
values  obtained  by  extrapolating  the on-orbit  $BVI$  sky  surface
brightness  of \citet{wind94,wind98} to  \zz, with  the sky-background
estimates from  the ACS  Instrument Handbook \citep{gonz05},  and with
the colors obtained by  convolving the filter transmission curves with
the  solar  spectrum.   \tabref{table1}  gives the  measured  electron
detection rate,  surface brightness and  colors of the  sky background
with  their  corresponding  errors  for  the HUDF  $BVi'z'$  bands  as
calculated from \figref{fig3}  and \figref{fig4}.  The contribution of
the zodiacal  background dominates the total  sky-background, which we
find to  be only $\sim$10\%  redder in (\vv--\ii) and  (\ii--\zz) than
the  Sun.   The  \zz-band  surface  brightness  corresponding  to  the
1$\sigma$ sky-subtraction uncertainty is therefore:
\begin{eqnarray*}
\mu_{z'} - 2.5\cdot \log (\Sigma_{\rm ss,ran}) &=& 22.577 - 2.5\cdot \log (2.97\times10^{-3}) \\
                                              &=& 28.895\ \hbox{mag arcsec}^{-2}
\end{eqnarray*}

Next, we measure the sky-background from 415 `\emph{blank}' sky stamps
(51$\times$51  pixel) distributed  throughout the  HUDF  (`global' sky
measurements).   We measure  the  sky background  using  the same  IDL
algorithm as used above.

\begin{figure}
\epsscale{1.0}
\plotone{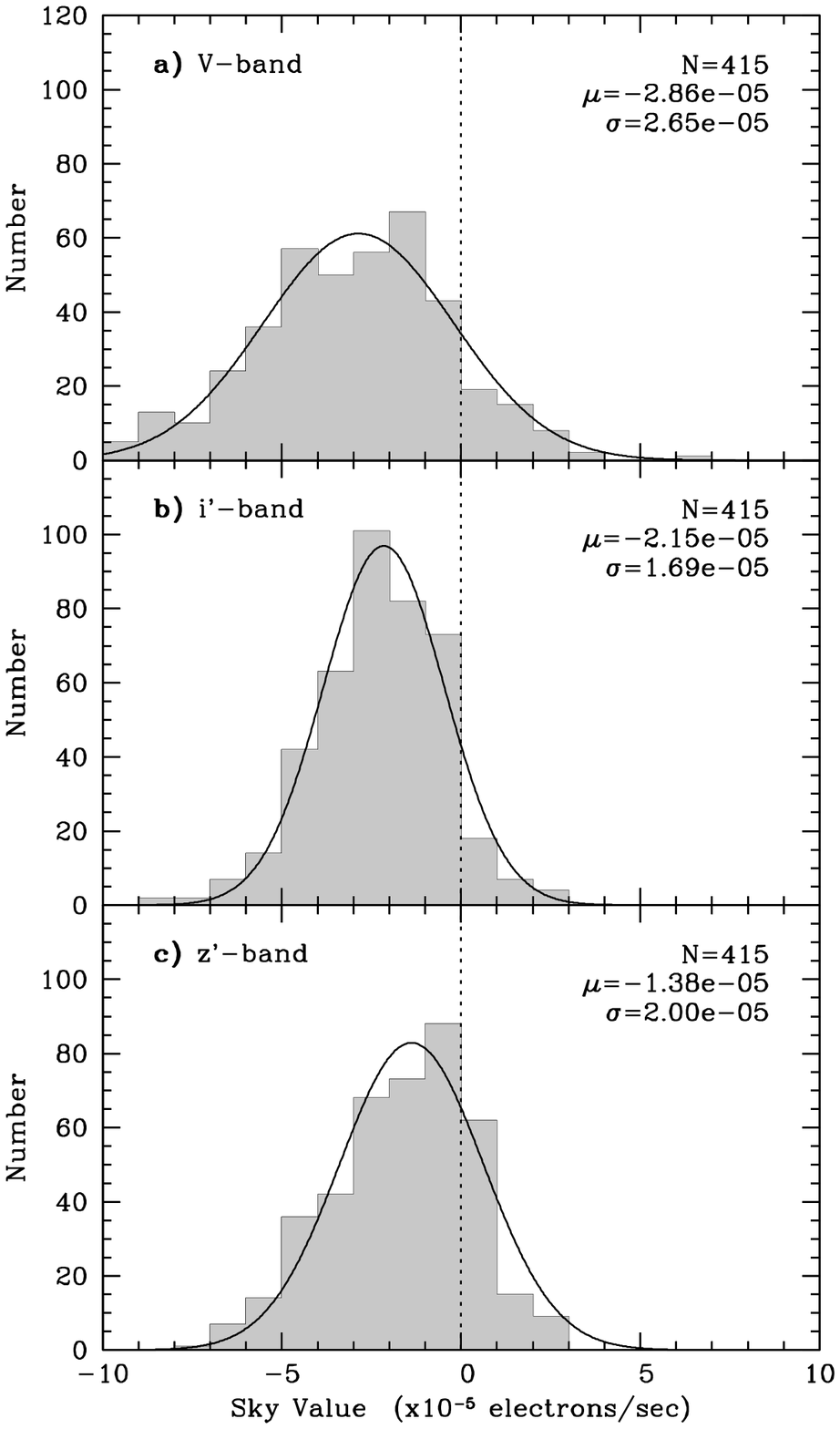}
\caption{Distribution  of  the  modal  sky background  level  used  to
estimate the 1$\sigma$  uncertainty in that level, as  measured in 415
\emph{`blank'  51$\times$51  pixel  sky  stamps}  extracted  from  the
drizzled HUDF  images {\bf (a)}  for \vv-band, {\bf (b)}  for \ii-band
and {\bf (c)} for \zz-band.  The mean ($\mu$) and the sigma ($\sigma$)
of to the best-fit Gaussian  to these distributions are shown in each
panel.}\label{fig5}
\end{figure}

\figref{fig5}c shows the histogram  of the measured \zz-band modal sky
values.  A Gaussian  distribution was fit to this  histogram, giving a
sky-sigma of $2.00\times10^{-5}$  $e^-$ sec$^{-1}$.  The average value
of the  sky remains 0.02051 $e^-$  sec$^{-1}$ (\figref{fig4}d).  Using
these   values,   we   can   estimate  a   relative   rms   systematic
sky-subtraction error as follows:
\[
\Sigma_{\rm ss,sys} = \frac{\sigma_{\rm sky,sys}}{{\rm I}_{\rm sky}} = \frac{2.00\times10^{-5}}{2.05\times10^{-2}\ \cdot\ 0.60^2} = 2.71\times10^{-3}
\]
Since  the   \zz-band  sky  surface  brightness   remains  22.577  mag
arcsec$^{-2}$, this gives us  for the surface brightness corresponding
to the 1$\sigma$ sky subtraction uncertainty:
\begin{eqnarray*}
\mu_{z'} - 2.5\cdot \log (\Sigma_{\rm ss,sys}) &=& 22.577 - 2.5\cdot \log (2.71\times10^{-3}) \\ 
                                               &=& 28.995\ \hbox{mag arcsec}^{-2}
\end{eqnarray*}

From  these two  complementary approaches,  we can  conclude  that all
surface     brightness    measurements    become     unreliable    for
surface-brightness    levels     fainter    than    28.95$\pm$0.05~mag
arcsec$^{-2}$ in the \zz-band. We have also experimented with slightly
larger cutouts (75$\times$75 pixels instead of 51$\times$51 pixels) to
estimate  the sky-subtraction  error.  We  find that  with  the larger
cutouts,  the  surface   brightness  corresponding  to  the  1$\sigma$
sky-subtraction  error is  $\sim$0.1--0.2  mag arcsec$^{-2}$  fainter.
For larger cutouts  we expect this surface brightness  to be $\sim$0.4
mag fainter  but we  find about 0.1--0.2  mag fainter.  This  might be
because of  residual systematic errors in the  HUDF images. Therefore,
we are  at the limit  of accurately measuring this  surface brightness
and hence, we  will here quote the conservative  brighter limit of the
surface  brightness corresponding  to  this 1$\sigma$  sky-subtraction
error.   Expected  contributions to  this  surface  brightness due  to
uncertainties in  the bias  level determinations, which  correspond to
$\sim$0.001  counts  sec$^{-1}$ for  typical  HUDF  exposures (A.   M.
Koekemoer; private communication), are less than 1\%.

\figref{fig5}  clearly  shows  that  the  distribution  of  the  modal
sky-values is  not as symmetric  around zero as in  \figref{fig3}, and
hence, the use of a `global' sky value for the HUDF is not as reliable
as `local'  sky measurements.   Therefore, for the  surface brightness
profiles and the following  discussion, we will adopt the \emph{local}
1$\sigma$ random sky-subtraction error for all objects in our study.  

The average  modal sky values and  their 1$\sigma$ errors  in the \vv-
and  \ii-bands were  calculated in  exactly the  same way  as  for the
\zz-band, as shown in \figref{fig3},  4 and 5.  The resulting $BVi'z'$
sky  values and  the sky  surface-brightness levels  are all  given in
\tabref{table1}.

\begin{deluxetable*}{cccccc}
\tabletypesize{\scriptsize}
\tablewidth{0pt}
\tablecaption{Measured sky values in $BVi'z'$ (filters) for the HUDF \label{table1}}
\tablenum{1}
\tablehead{\colhead{HUDF} & \colhead{Number of} & \colhead{Mean Sky Value$^a$} & \colhead{Sky SB$^c$} & 
\colhead{Sky Color$^c$} & \colhead{1$\sigma$ Sky-Subtraction} \\
\colhead{Filter} & \colhead{Exposures} & \colhead{($e^-$/s) and rms error$^b$} & \colhead{(AB mag arcsec$^{-\
2}$)} & \colhead{(AB mag)} & \colhead{error (AB mag arcsec$^{-2}$)} }

\startdata
$B$ & 112 & 0.015909 $\pm$ 0.000065 & 23.664 $\pm$ 0.003 & ($B-V$)$_{\rm sky}$=0.800 & 29.85 $\pm$ 0.05 \\
$V$ & 112 & 0.070276 $\pm$ 0.000297 & 22.864 $\pm$ 0.002 & ($V-i'$)$_{\rm sky}$=0.222 & 30.15 $\pm$ 0.15 \\
$i'$ & 288 & 0.040075 $\pm$ 0.000088 & 22.642 $\pm$ 0.002 & ($i'-z'$)$_{\rm sky}$=0.065 & 29.77 $\pm$ 0.20 \\
$z'$ & 288 & 0.020511 $\pm$ 0.000047 & 22.577 $\pm$ 0.003 & ($V-z'$)$_{\rm sky}$=0.287 & 28.95 $\pm$ 0.05 \\
\enddata

\tablenotetext{a}{From \figref{fig4}}
\tablenotetext{b}{Error is standard deviation of the mean ($\sigma$/$\surd{N}$)}
\tablenotetext{c}{Sky surface brightness values and colors are consistent with the solar colors in AB mag of
(\vv-\ii)=0.19, (\vv-\zz)=0.21 and (\ii-\zz)=0.01 [except for bluest color (\bb-\vv)], and is dominated by
the zodiacal background.}
\end{deluxetable*}

\subsection{Composite Images and Surface Brightness Profiles}

\begin{figure}
\epsscale{1.0}
\plotone{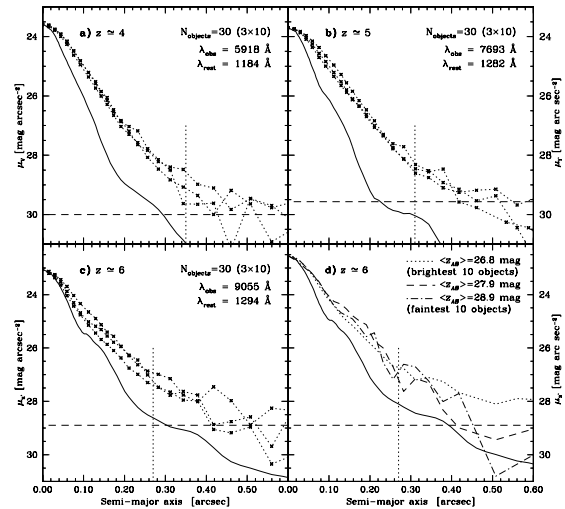}
\caption{Composite surface  brightness profiles for  three independent
sets   of  10  objects   at  {\bf   (a)}  $z\!\simeq\!4$,   {\bf  (b)}
$z\!\simeq\!5$ and  {\bf (c)} $z\!\simeq\!6$,  respectively.  The thin
solid curve  represents the ACS \vv, \ii\thinspace  and \zz-band PSFs,
respectively, while  the horizontal dashed line  indicates the surface
brightness level corresponding to the 1$\sigma$ sky--subtraction error
in the HUDF images. The vertical dotted line marks the radius at which
the profile starts to  deviate significantly from the extrapolation of
the inner  $r^{1/n}$ profile observed  at smaller radii. Note  that at
$z\!\simeq\!6$, this  deviation is still  well above the  red \zz-band
PSF  halo   at  $r$\cge0\farcs30.   The   panel  {\bf  (d)}   shows  3
$z\!\simeq\!6$  composite profiles  (each with  a set  of  10 objects)
divided  by  apparent  magnitudes.   The brightest  composite  profile
(dotted)  has an average  \zz-band magnitude  of $\sim$26.8  mag.  The
next brightest composite profile  (short dash) has an average \zz-band
magnitude  of  $\sim$27.9  mag,  and the  faintest  composite  profile
(dot-dash)   has   an  average   \zz-band   magnitude  of   $\sim$28.9
mag.}\label{fig6}
\end{figure}

For  each  redshift   bin  ($z\!\simeq\!4,5,6$),  we  generated  three
``stacked'' composite  images from subsets  of 10 postage  stamps that
were  selected as  follows.  After  placing  all 30  image stamps  per
redshift  bin into  a $30\times$  (51$\times$51) pixel  IDL  array, 10
stamps  were randomly  drawn without  selecting any  object  more than
once.   An output image  was generated,  in which  the values  at each
pixel are the  average of the corresponding pixels  in the 10 selected
input stamps. From  the remaining 20 stamps, we  again randomly select
10, from which we generated  a second composite image, after which the
final 10  images were  averaged into the  third composite  image.  The
three composite  images per redshift bin are  therefore independent of
each  other.  In  none of  our realizations  did we  produce composite
images   that  were   essentially  unresolved.    Even   the  faintest
$z\!\simeq\!4\!-\!6$    galaxies    are    clearly   resolved.     The
$z\!\simeq\!4,5,6$ objects used to  generate the composite images have
an  apparent magnitude  range  of approximately  27.5$\pm$1.0 AB  mag.
Because the magnitude range is  relatively small and the S/N per pixel
is low  even in their central  pixel, we have given  all objects equal
weight.   To test  whether this  range  in magnitude  will affect  our
stacks and hence,  our profiles, we created 3  stacks depending on the
apparent magnitude, i.e. one stack  of the 10 brightest objects in the
sample, a second stack of the  10 next brightest objects in the sample
and a  third stack of the 10  faintest objects in the  sample. This is
summarized  in \figref{fig6}d. We  found that  the profiles  were very
similar except that  the profiles of the fainter  stacks fall-off more
quickly  at larger  radius compared  to the  profile of  the brightest
stack, but  the inner  profile and the  deviation in the  profiles are
clearly visible in all 3  stacks.  Therefore, we conclude that for our
range   in   apparent   magnitudes,   our  stacks/profiles   are   not
affected. Perhaps  most surprisingly, \figref{fig6}d  shows that $r_e$
value  of all  3  flux ranges  ($\sim$26.0--27.0, $\sim$27.0--28.0  \&
$\sim$28.0--29.0 mag)  are all  about the same  over $\sim$3-4  mag in
flux, so  the primary parameter  that distinguishes the  brighter from
the   fainter  $z\!\simeq\!6$  dropouts   is  their   central  surface
brightness (which thus also varies by $\sim$3--4 mag).

\begin{figure}
\epsscale{1.0}
\plotone{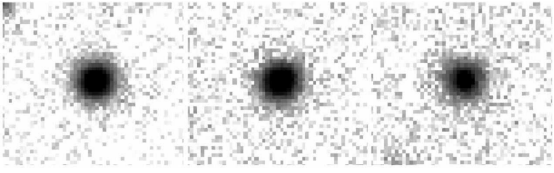}
\caption{Composite images for  {\bf Left} $z\!\simeq\!4$, {\bf Center}
$z\!\simeq\!5$  and  {\bf Right}  $z\!\simeq\!6$  objects.  Here  each
composite image is a stack of 30 objects. Each stamp is $1\farcs53$ on
a side.}\label{fig7}
\end{figure}

\begin{figure}
\epsscale{1.1}
\plottwo{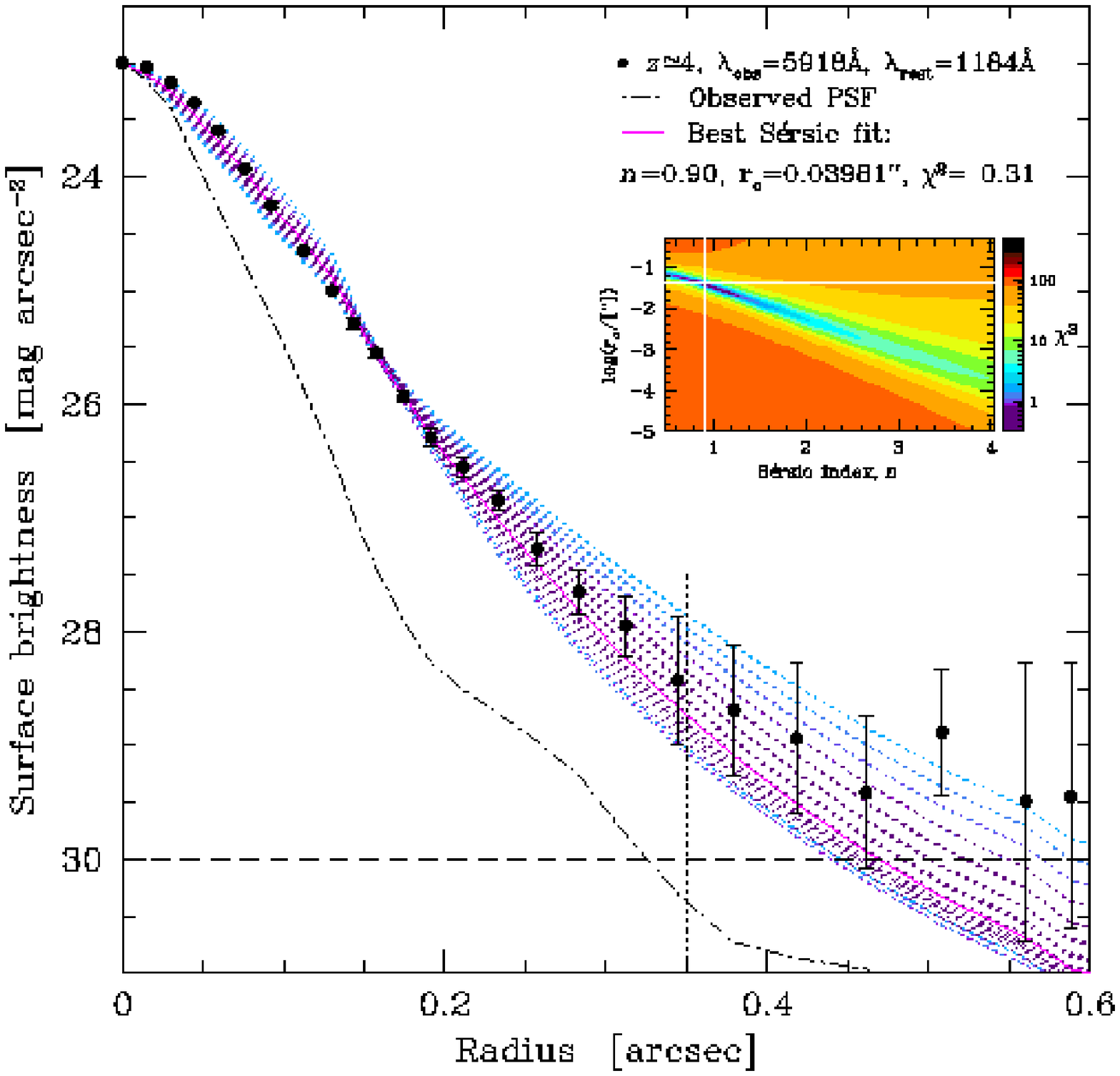}{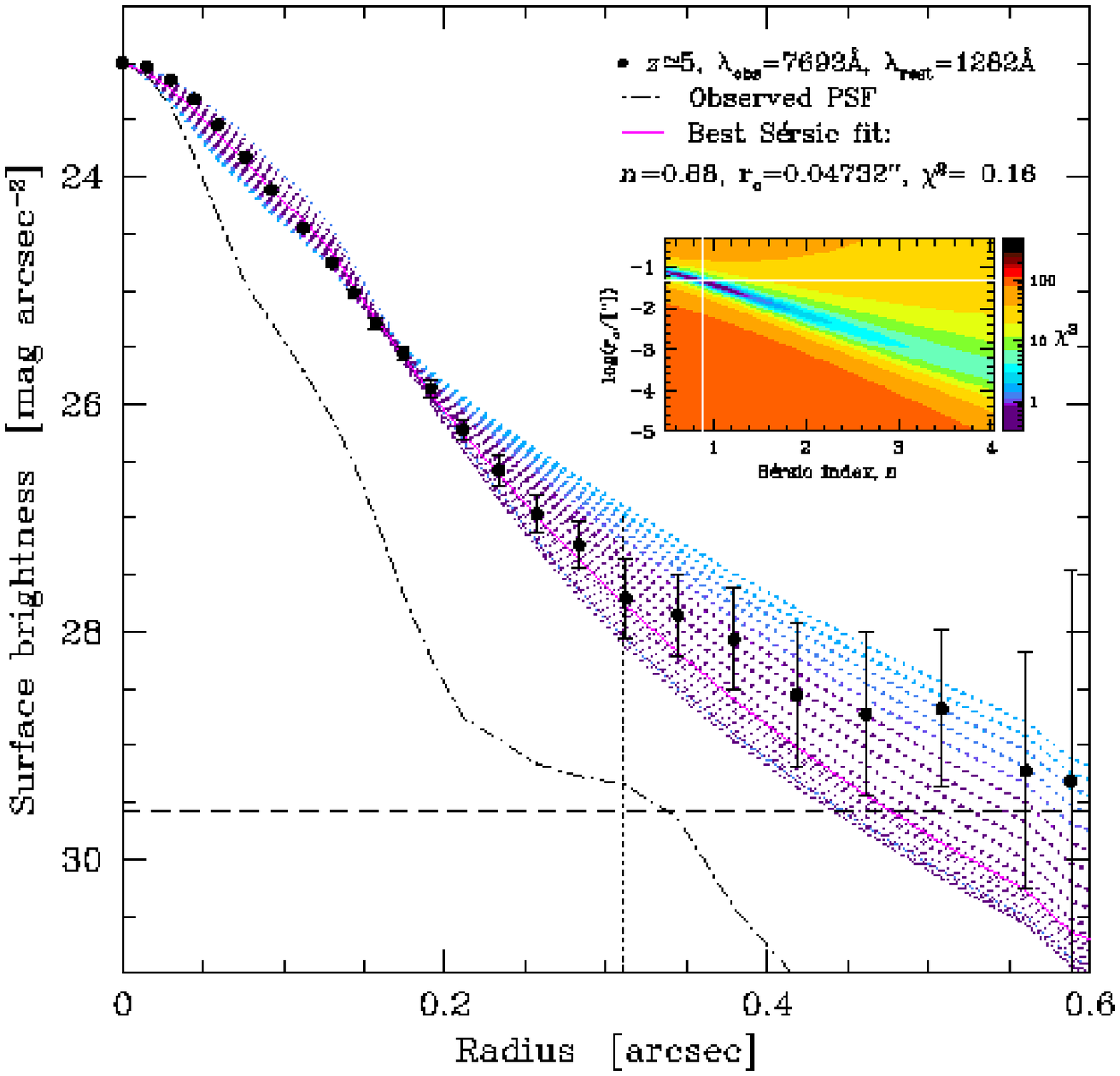}
\epsscale{0.5}
\plotone{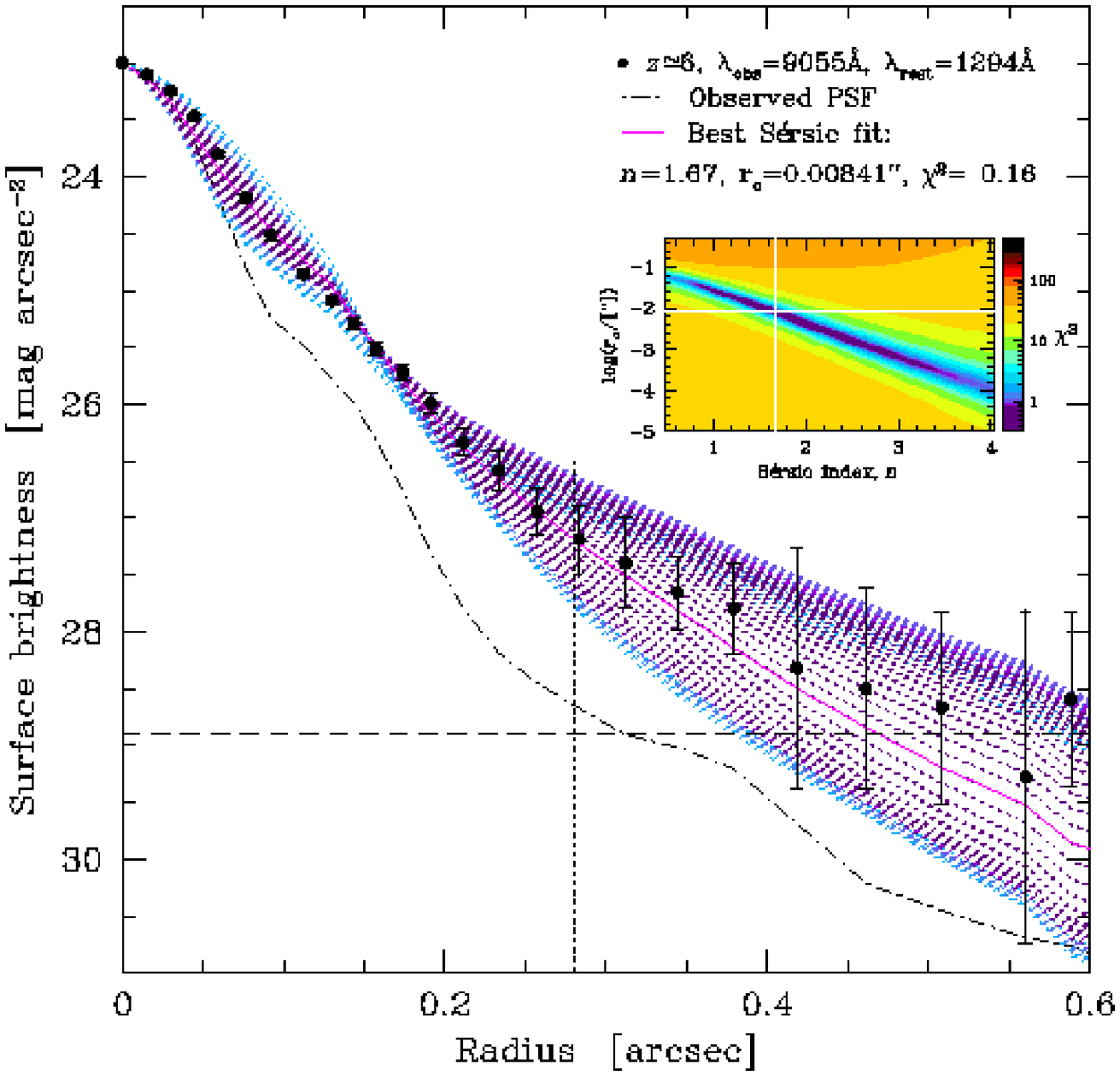}
\caption{Mean surface  brightness profiles with a  best fit S\'{e}rsic
profiles  for 30 composite  images at  {\bf (a)}  $z\!\simeq\!4$, {\bf 
(b)} $z\!\simeq\!5$  and {\bf (c)}  $z\!\simeq\!6$, respectively.  The
thin dot-dash curve represents the ACS \vv, \ii\thinspace and \zz-band
PSFs,  respectively, while  the horizontal  dashed line  indicates the
surface    brightness   level    corresponding   to    the   1$\sigma$
sky--subtraction error  in the HUDF  images. The vertical  dotted line
marks the radius at which  the profile starts to deviate significantly
from  the extrapolation  of the  inner $r^{1/n}$  profile  observed at
smaller  radii.  The  $n$  and  r$_c$  are  the  best  fit  S\'{e}rsic
parameters.  }\label{fig8}
\end{figure}

We used the IRAF\footnote{IRAF (http://iraf.net) is distributed by the
National Optical  Astronomy Observatories,  which are operated  by the
Association  of Universities  for Research  in Astronomy,  Inc., under
cooperative   agreement  with   the   National  Science   Foundation.}
procedure \texttt{ELLIPSE} to fit surface brightness profiles shown in
\figref{fig6} to  each of the  three independent composite  images per
redshift bin.  We also computed a mean surface-brightness profile from
the  three composite  surface brightness  profiles generated  from the
three   independent   composite   images   for  each   redshift   bin.
\figref{fig7} shows  composite images for  $z\!\simeq\!4,5,6$ objects.
Here each  composite image  is a stack  of 30  objects.  
\figref{fig8} shows  the average surface brightness  profiles for each
of the  redshift intervals $z\!\simeq\!4,5,6$.  The  thin solid curves
in \figref{fig6}  and the  dot-dash curves in  \figref{fig8} represent
the  observed  ACS  \vv,   \ii\thinspace  and  \zz-band  Point  Spread
Functions  (PSFs),  while the  horizontal  dashed  lines indicate  the
surface    brightness   level    corresponding   to    the   1$\sigma$
sky--subtraction  error in  each of  the HUDF  images as  discussed in
\secref{skyerror}.  It  is important  to note that  we scaled  the ACS
PSFs to match the surface brightness  of the central data point in our
mean surface-brightness  profile, to  determine how extended  the mean
surface-brightness profile is with respect to the PSFs.

In  \figref{fig8},   we  fitted  all  possible   combinations  of  the
Sers\'{i}c  profiles (convolved  with  the ACS  PSF)  to the  observed
profiles  and using  $\chi^2$ minimization,  found the  best  fits for
galaxies at  $z\!\simeq\!4,5,6$.  The best fit  S\'{e}rsic index ($n$)
for  all  three profiles  ($z\!\simeq\!4,5,6$)  is $n\!<\!2$,  meaning
these galaxies  follow mostly exponential disk-type  profiles in their
central regions.  We find that  the observed profiles start to deviate
from  the  best-fit  profiles  at  $r\!\gtrsim$0\arcspt  27,  somewhat
depending on  the redshift.  From  \figref{fig8}, we also see  that in
each   of  $V$  ($z\!\simeq\!4$),   $i'$  ($z\!\simeq\!5$)   and  $z'$
($z\!\simeq\!6$), the  PSF declines more rapidly with  radius than the
composite radial  surface brightness profile  for $r\!\gtrsim$0\arcspt
27.  It is  therefore unlikely that the observed  `breaks' result from
the  halos   and  structure  of   the  ACS  PSFs.    Specifically,  at
$z\!\simeq\!6$ the  most significant deviations  in the light-profiles
are seen  at levels 1.5--2.0  mag above the  1$\sigma$ sky-subtraction
error,  and  well above  the  PSF wings.   Each  of  the mean  surface
brightness profiles display a  well-defined break, the radius of which
appears to  change somewhat with  redshift. The vertical  dotted lines
(in \figref{fig6} and \figref{fig8}) mark the radius at which the mean
surface brightness  profiles start  to deviate significantly  from the
extrapolation of the $r^{1/n}$ profile observed at smaller radii.


\subsection{Test of the Stacking Technique on Nearby Galaxies}

To test  the general  validity of the  stacking technique itself  on a
local galaxy sample, we used  surface photometry from the Nearby Field
Galaxy  Survey   \citep[NFGS:][]{jans00a,jans00b}.   The  NFGS  sample
contains 196 nearby galaxies,  that were objectively selected from the
CfA redshift catalog \citep[CfA\,I;][]{davi83,huch83} to span the full
range in absolute $B$  magnitude present in the CfA\,I ($-14.7\lesssim
\!M_B\!\lesssim -22.7$  mag).  The absolute  magnitude distribution in
the  NFGS sample  approximates  the local  galaxy luminosity  function
\citep[e.g.,][]{marz94},  while  the  distribution  over  Hubble  type
follows  the changing  mix of  morphological  types as  a function  of
luminosity in the local  galaxy population.  The NFGS sample \citep[as
detailed in][]{jans00a}  minimizes biases,  and yields a  sample that,
with  very  few  caveats,   is  representative  of  the  local  galaxy
population.   As part  of  the NFGS,  $UBR$  surface photometry,  both
integrated (global) and nuclear spectrophotometry, as well as internal
kinematics  were   obtained  \citep[see][]{jans01}.   Here,   we  will
concentrate on the $U$-band surface photometry, since it is closest in
wavelength    to    the    rest-frame    wavelengths    observed    at
$z\!\simeq\!4\!-\!6$.   Although,  ideally,  we  would want  a  filter
further into  the UV, \citet{tayl07} and \citet{wind02}  show that for
the majority  of late-type nearby galaxies, the  apparent structure of
galaxies does  not change dramatically once one  observes shortward of
the Balmer break. Early-type  galaxies, however, are a clear exception
to this, but these are  not believed to dominate the galaxy population
at $z\!\simeq\!4\!-\!6$, as discussed before.

\begin{figure}
\epsscale{1.0}
\plotone{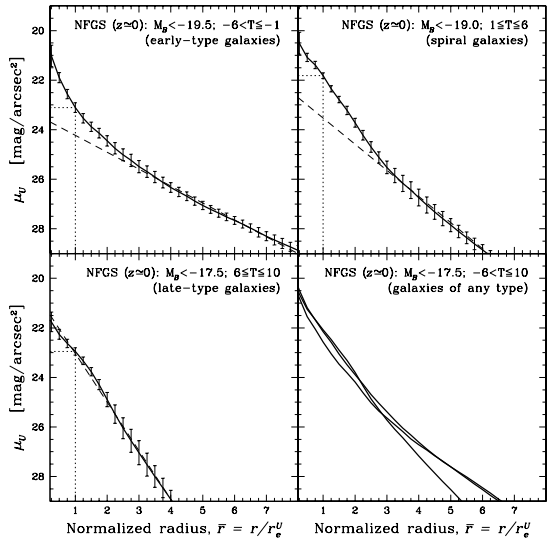}
\caption{Stacked  radial surface  brightness  profiles for  relatively
luminous early-, spiral- and  late-type nearby galaxies drawn from the
Nearby  Field  Galaxy  Survey \citep{jans00a,jans00b}.   The  vertical
dotted  line indicates the  half-light radius,  while the  dashed line
represents an exponential fit to  the outer portions of each composite
profile.  Co-adding profiles for disparate morphological types and for
spiral  galaxies with  a  range in  bulge-to-disk  ratios can  produce
breaks in the composite profile. No significant breaks are seen in the
\emph{outer} light  profiles, when  the profiles of  either early-type
galaxies    (E,   S0)    or   late-type    galaxies    (Sd--Irr)   are
co-added.}\label{fig9}
\end{figure}

\figref{fig9} shows  stacked profiles for  relatively luminous early-,
spiral-, and  late-type galaxies drawn from the  NFGS. Vertical dotted
lines indicate  the half-light radii  and their intersection  with the
profiles,  the  surface  brightness  at  that  radius.   Dashed  lines
indicate  exponential  fits to  the  outer  portion  of each  profile.
\figref{fig9}  also  shows   that  co-adding  profiles  for  disparate
morphological types and  for mid-type spiral galaxies with  a range in
bulge-to-disk ratios  can produce breaks in the  composite profile. No
such breaks are  seen when the profiles of  either early-type galaxies
(E,  S0) or late-type  galaxies (Sd--Irr)  are co-added.   This figure
shows  that, \emph{if}  galaxies at  $z\!\simeq\!4\!-\!6$  had similar
morphological types  as local galaxies,  then it would be  possible to
produce  a  break in  the  profiles  (as  shown in  \figref{fig6}  and
\figref{fig8}), merely  by mixing different types of  galaxies.  We do
not  believe  that  the  galaxy  populations  at  $z\!\simeq\!4\!-\!6$
morphologically resemble those at  low redshift.  Hence, for primarily
late-type galaxies, which dominate the faint blue galaxy population at
AB$\ge$24 mag  \citep{driv98}, and  which likely dominate  the fainter
end of the luminosity  function at $z\!\simeq\!4\!-\!6$ that we sample
here  \citep{yan04a,yan04b},  the image  stacking  is  likely a  valid
exercise.

The primary goal of this section was to show that the profile stacking
technique  is  valid  and  can  be  used  to  get  meaningful  surface
brightness  profiles.  We  are not  comparing our  nearby  sample with
galaxies at  $z\!\simeq\!4\!-\!6$. These nearby  galaxies are unlikely
to be local analogues of  high redshift galaxies.  If we apply surface
brightness dimming to UV light-profiles of these nearby galaxies, they
would be mostly invisible to  \emph{HST}, and in some cases visible to
\emph{JWST} in  long integration \citep[see  e.g.][]{wind06}.  This is
another way of saying  that the $z\!\simeq\!4\!-\!6$ objects are truly
different from $z\!\simeq\!0$ objects.


\section{Discussions}\label{results}

\figref{fig8} shows that the  mean surface brightness profiles deviate
significantly   from    an   inner   $r^{1/n}$    profile   at   radii
$r\!\gtrsim$0\arcspt  27--0\arcspt  35,   depending  somewhat  on  the
redshift bin.   These deviations appear real, with  the break/point of
departure located \cge  1.5--2~mag above the 1$\sigma$ sky-subtraction
error and above  the PSF-wings.  In the following,  we discuss several
possible explanations for the observed shapes of our composite surface
brightness profiles.

\subsection{Galaxies with Different Morphologies}

Our test  on nearby galaxies  (\figref{fig9}) shows that, if  we stack
many  galaxies with different  morphologies (early-type,  late-type or
spiral galaxies),  it is possible  to get a slope-change  (`break') in
the average surface brightness  profile. \citet{ravi06} find that 40\%
of the brighter LBGs at $2.5\!<\!z\!<\!5$ have light profiles close to
exponential, as seen for disk  galaxies, and only $\sim$30\% have high
$n$,  as seen  in  nearby  spheroids.  They  also  find a  significant
fraction ($\sim$30\%) of galaxies with light profiles \emph{shallower}
than  exponential, which appear  to have  multiple cores  or disturbed
morphologies, suggestive  of close  pairs or on-going  galaxy mergers.
Therefore,  if  galaxies at  $z\!\simeq\!4\!-\!6$  have  a variety  of
morphological types, then the  shape of the average surface brightness
profile that we  see may be due to the stacking  of different types of
galaxies.   Therefore, we find  that the  exponential and  the flatter
profiles  found by  \citet{ravi06} for  galaxies  at $2.5\!<\!z\!<\!5$
also apply to higher redshifts ($z\!\ge\!5$).

Also, we believe that it is  more likely that the high redshift, faint
galaxy population consists primarily  of small galaxies with late-type
morphologies   and   with  sub-L$^{*}$   luminosities,   as  seen   at
$z\!\simeq\!2\!-\!3$     \citep{driv95,driv98}.      So     if     the
$z\!\simeq\!4\!-\!6$  population consists of  such a  late-type galaxy
population, then the slope-change in  the light profiles is likely not
the result of co-adding images of objects with disparate morphological
types.

\subsection{Central Star Formation/Starburst}

\emph{HST} optical  images of galaxies  at $z\!\simeq\!4\!-\!6$ sample
their rest-frame UV ($\sim$1200 \Ang), where the contribution from the
actively  star-forming regions (very  young, massive  stars) dominates
the   UV-light.    \citet{hath07}   have   shown  that   galaxies   at
$z\!\simeq\!5\!-\!6$ are  high redshift starbursts  and these galaxies
have similar starburst intensity limit as local starbursting galaxies.
Therefore, it  is possible that galaxies  at $z\!\simeq\!4\!-\!6$ have
centrally concentrated star  formation or starburst.  This possibility
is  based  on three  key  assumptions: (1)  most  of  the galaxies  at
$z\!\simeq\!4,5,6$     are    intrinsically     later-type    galaxies
\citep{driv98,stei99}; (2)  the Spectral Energy  Distribution (SED) of
these galaxies at $z\!\simeq\!4,5,6$ are dominated by early A- to late
O-type stars, respectively;  and (3) there are no  old stars with ages
at  $z\!\simeq\!4\!-\!6$  greater  than  2-1 Gyr  in  WMAP  cosmology,
respectively.

\citet{hunt06}   studied  azimuthally   averaged   surface  photometry
profiles for large sample of nearby irregular galaxies. They find some
galaxies have double exponentials that  are steeper (and bluer) in the
inner  parts compared to  outer parts  of the  galaxy.  \citet{hunt06}
discuss that this  type of behavior is expected  in galaxies where the
centrally  concentrated  star  formation  or  starburst  steepens  the
surface brightness profiles  in the center. If that  is the case, then
one might expect a better correlation between the break in the surface
brightness profiles and changes  in color profiles. Unfortunately, for
our sample  of galaxies at  $z\!\simeq\!4\!-\!6$, we don't  have high-
resolution  restframe \uu\bb\vv  color information.   The  objects are
generally too  faint for \emph{Spitzer Space Telescope},  and hence we
cannot  confirm  or reject  this  possibility  for  the shape  of  our
composite surface brightness profiles.

\subsection{\boldmath {Limits to Dynamical Ages for $z\!\simeq\!4,5,6$ Objects}}\label{ages}

The  average compact $z\!\simeq\!4\!-\!6$  galaxy is  clearly extended
with respect  to the ACS PSFs  (\figref{fig8}), and is best  fit by an
exponential   profile   ($n\!<\!2$)  out   to   a   radius  of   about
$r\!\simeq$0\arcspt   35,   0\arcspt    31,   and   0\arcspt   27   at
$z\!\simeq\!4,5$ and $6$, respectively.  The apparent progression with
redshift is  noteworthy.  The  radius at which  the profile  starts to
deviate  from  $r^{1/n}$ (in  this  case  at  radius $r$\cge  0\arcspt
35--0\arcspt 27) may be an  important constraint to the dynamical time
scale of the system,  as discussed in \secref{introduction}.  If this
argument is valid,  then we can estimate limits  to the dynamical ages
of $z\!\simeq\!4,5,6$ galaxies as follows.
  
In WMAP cosmology,  a radius of $r$\cge 0\arcspt  35 at $z\!\simeq\!4$
corresponds   to  $r$\cge   2.5   kpc.   The   dynamical  time   scale
\citep[e.g.,][]{binn87},  $\tau_{dyn}$,  goes  as  $\tau_{dyn}$  =  $C
r^{3/2} \!/\!\sqrt{G\,M}$, where the constant $C=\pi/2$. For a typical
dwarf  galaxy mass  range of  $\sim10^9\!-\!10^8$\Msun  inside $r$=2.5
kpc,  we  infer  that  the  limits  to  the  dynamical  age  would  be
$\tau_{dyn}\simeq$ 90--290  Myr, which is the lifespan  expected for a
late-type B-star.  This means that the last major merger that affected
this  surface brightness  profile  and that  triggered its  associated
starburst  may  have occurred  $\sim$0.20  Gyr before  $z\!\simeq\!4$,
---assuming that the star-formation wasn't spontaneous, but associated
with some accretion or a merging event.

Table 2 shows  the break-radius and inferred limits  to dynamical ages
for the $z\!\simeq\!4\!-\!6$ objects.  At $z\!\simeq\!5$, we find that
the  limits   to  dynamical   age  at  the   break  radius   would  be
$\tau_{dyn}\simeq$ 70--210  Myr, which is the lifespan  expected for a
mid B-star,  while at $z\!\simeq\!6$,  $\tau_{dyn}\simeq$ 50--150 Myr,
which is the lifespan expected for a late O--early B-star.  This means
that  the last  major merger  that affected  these  surface brightness
profiles at  $z\!\simeq\!5$ and $6$ and that  triggered its associated
starburst  may  have occurred  $\sim$0.14  and  $\sim$0.10 Gyr  before
$z\!\simeq\!5$ and $6$, respectively.

\begin{deluxetable}{cccc}
\tablewidth{0pt}
\tablecaption{Dynamical Ages for $z\!\simeq\!4-6$ objects in the HUDF\label{table2}}
\tablenum{2}
\tablehead{\colhead{Redshift} & \colhead{``Break'' Radius$^a$} & \colhead{``Break'' Radius$^b$} & 
\colhead{Dynamical Age$^c$} \\
\colhead{$z$} &      \colhead{(arcsec)} &   \colhead{(kpc)} &      \colhead{($\tau_{dyn}$)} }

\startdata
4 & 0.35 & 2.5 & 0.09--0.29 Gyr \\
5 & 0.31 & 2.0 & 0.07--0.21 Gyr \\
6 & 0.27 & 1.6 & 0.05--0.15 Gyr \\
\enddata

\tablenotetext{a}{From composite surface brightness profiles (\figref{fig6} and \figref{fig8}).}
\tablenotetext{b}{Radius in kpc corresponding to radius in arcsec at given redshift.}
\tablenotetext{c}{If ``break radius'' interpreted as indicator of dynamical age.}
\end{deluxetable}

The  dynamical time is  a lower  limit to  the actual  time available,
since it  assumes matter  starts from rest.   Any angular  momentum at
start will increase the available time.  The best-fit SED age from the
GOODS \emph{HST} and \emph{Spitzer} photometry on some of the brighter
of  these objects  --- using  \citet{bruz03} templates  --- is  in the
range  of about  $\sim$150--650  Myr \citep{yan05,eyle05,eyle07},  the
lower end  of which is consistent  with our limits  to their dynamical
age  estimates, while  the  somewhat  larger SED  ages  could also  be
affected by the onset of  the AGB in the stellar population increasing
the observed  \emph{Spitzer} fluxes and  hence possibly overestimating
ages \citep{mara05}.   Our age estimates  for $z\!\simeq\!4\!-\!6$ are
consistent  with the trend  of SED  ages suggested  for $z\!\simeq\!7$
\citep{labb06}.  It  is noteworthy that, given  the uncertainties, the
two  independent  age  estimates  are  consistent. If  our  limits  to
dynamical age  estimates for the  image \emph{stacks} are  thus valid,
they are consistent with the SED ages, and point to a consistent young
age for these objects.

Furthermore, the  presence of young, massive late  O--early B-stars at
$z\!\simeq\!6$ has implications for  the reionization of the universe.
From observations of the  appearance of complete Gunn-Peterson troughs
in the  spectra of $z\!\ga\!5.8$  quasars \citep{fan06}, we  know that
the epoch of reionization had ended by $z\!\simeq\!6$.  From the steep
($\alpha$=--1.8)  faint-end  slope   of  the  luminosity  function  of
$z\!\simeq\!6$  galaxies, \citet{yan04a,yan04b}  concluded  that dwarf
galaxies,   and   not  quasars,   likely   finished  reionization   by
$z\!\simeq\!6$.   Should  the present  interpretation  of their  light
profiles  be correct,  then it  would appear  to add  support  to this
picture, in the  sense that such objects are  dominated by B-stars and
did  not   start  their  most  recent  major   starburst  long  before
$z\!\simeq\!6$.

\section{Summary}\label{conclusion}

We  used  the stacked  HUDF  images  to  analyze the  average  surface
brightness  profiles of  $z\!\simeq\!4\!-\!6$ galaxies.   Our analysis
shows  that even  the  faintest galaxies  at $z\!\simeq\!4\!-\!6$  are
resolved. This  may have implications  on the stellar density  and its
relation to the stellar density in present-day galaxies.  We also find
that  the average  surface  brightness profiles  display  breaks at  a
radius that progresses toward lower redshift from $r\simeq$0\arcspt 27
(1.6  kpc) at  $z\!\simeq\!6$  to $r\simeq$0\arcspt  35  (2.5 kpc)  at
$z\!\simeq\!4$.

The shape  of the  radial surface brightness  profile that  we observe
could  result  from a  mixture  of  different  morphological types  of
galaxies,  if  they  exist  at $z\!\simeq\!4\!-\!6$,  because  we  can
produce similar breaks in the  surface brightness profiles when we mix
different types of nearby  galaxies.  Alternatively, if these galaxies
are dominated by a central  starburst then they could show such double
exponential-type profiles, as discussed by \citet{hunt06}.  In a third
scenario,  if the  galaxies at  $z\!\simeq\!4-6$ are  truly  young and
mostly late-type, the  outer profiles seen in our  mean radial surface
brightness  profiles  at  $z\!\simeq\!4-6$  bear the  imprint  of  the
hierarchical build-up  process, and  are still dominated  by infalling
material, which is \emph{not} detectable in the individual HUDF images
of these  faint objects.  We  have estimated limits to  dynamical ages
from  the  break radius  at  $z\!\simeq\!4,  5,  6$, very  roughly  as
$\sim$0.20,  0.14  and 0.10  Gyr,  respectively,  and  those ages  are
similar   to   the   SED   ages   inferred   at   $z\!\simeq\!4\!-\!6$
\citep{yan05,eyle05,eyle07},  and consistent  with SED  ages suggested
for  $z\!\simeq\!7$ \citep{labb06}.  Hence,  at $z\!\simeq\!4,  5, 6$,
the last  major merger that  affected the surface  brightness profiles
that we observe, and that  triggered the observed star-burst, may have
occurred respectively  $\sim$0.20, 0.14 and 0.10 Gyr  earlier, or very
approximately at $z\!\simeq\!4.5, 5.5, 6.5$.  This would be consistent
with  the  hierarchical assembly  of  galaxies  and  with the  end  of
reionization,  since  it  would  imply  that  from  $z\!\simeq\!4$  to
$z\!\simeq\!6$,  the  SEDs  become  progressively  more  dominated  by
late-B--late-O stars.   This implies that the  sub-$L^*$ (i.e.  dwarf)
galaxies may have produced  sufficient numbers of energetic UV photons
to   complete   the  reionization   process   by  $z\!\simeq\!6$,   as
\citet{yan04a,yan04b} suggested.  It will  be imperative to study with
future    instruments    like    \emph{HST}/WFC3    and    \emph{JWST}
\citep{wind06,wind07} whether  the dominant stellar  population indeed
changes  from  late-O--early-B  at  $z\!\simeq\!6$ (i.e.   capable  of
reionizing) to  mid- to late-B  at $z\!\simeq\!4-5$ (i.e.   capable of
maintaining reionization),  and to what extent the  intrinsic sizes of
these faint objects will ultimately limit deep \emph{JWST} surveys.


\acknowledgments This  work was partially  supported by HST  grants AR
10298 and GO 9780 from the Space Telescope Science Institute, which is
operated by AURA  under NASA contract NAS 5-26555.   The authors thank
Deidre Hunter, Alan Dressler,  Henry Ferguson, Anton Koekemoer, Robert
Morgan for  their helpful  discussions. RAW acknowledges  support from
NASA JWST Interdisciplinary Scientist grant NAG5-12460 from GSFC, that
supported an investigation of the  implications of this work for JWST.
We specially thank our referee, Dr.  Patrick McCarthy, for his helpful
comments that have improved this paper.

Facilities: \facility{HST(ACS)}



\begin{thebibliography}{}
\bibitem[Beckwith  \etal(2006)]{beck06} Beckwith,  S.,  Stiavelli, M.,
Koekemoer, A. M., et al. 2006, AJ, 132, 1729
\bibitem[Bertin  \&  Arnouts(1996)]{bert96}  Bertin, E.,  \&  Arnouts,
S. 1996, A\&AS, 117, 393
\bibitem[Binney \& Tremaine(1987)]{binn87} Binney, J. J., \& Tremaine,
S. 1987, Galactic Dynamics (Princeton: Princeton Univ. Press)
\bibitem[Bolzonella  \etal(2000)]{bolz00}  Bolzonella,  M.,  Miralles,
J. M., \& Pell\'{o}, R. 2000, A\&A, 363, 476
\bibitem[Bouwens \etal(2004)]{bouw04} Bouwens, R., Illingworth, G. D.,
Thompson, R. I., et al. 2004, ApJ, 606, L25
\bibitem[Bouwens  \etal(2006)]{bouw06}  Bouwens,  R. J.,  Illingworth,
G.  D.,  Blakeslee,   J.  P.,  \&  Franx,  M.   2006,  ApJ,  653, 53
\bibitem[Bouwens  \etal(2007)]{bouw07}  Bouwens,  R. J.,  Illingworth,
G. D., Franx, M., \& Ford, H. 2007, ApJ, in press (astro-ph/0707.2080)
\bibitem[Brandt  \etal(2001)]{bran01}  Brandt,  W. N.,  Hornschemeier,
A.  E., Schneider, D.  P., Alexander,  D. M.,  Bauer, F.  E., Garmire,
G. P., \& Vignali, C. 2001, ApJ, 558, L5
\bibitem[Bruzual  \& Charlot(2003)]{bruz03}  Bruzual, G.,  \& Charlot,
S. 2003, MNRAS, 344, 1000
\bibitem[Bunker \&  Stanway(2004)]{bunk04} Bunker, A.  J., \& Stanway,
E. R. 2004, (astro-ph/0407562)
\bibitem[Cowie  \etal(1996)]{cowi96} Cowie, L.  L., Songaila,  A., Hu,
E. M., \& Cohen, J. G. 1996, AJ, 112, 839
\bibitem[Davis  \&  Peebles(1983)]{davi83}   Davis,  M.,  \&  Peebles,
P. J. M. 1983, ApJ, 267, 465
\bibitem[Dow-Hygelund     \etal(2007)]{dow07}     Dow-Hygelund,    C.,
Holden,  B., Bouwens, R., et al. 2007, ApJ, 660, 47
\bibitem[Driver \etal(1995)]{driv95} Driver,  S. P., Windhorst, R. A.,
\& Griffiths, R. E. 1995, ApJ, 453, 48
\bibitem[Driver  \etal(1998)]{driv98} Driver,  S.  P., Fernandez-Soto,
A.,  Couch, W. J.,  Odewahn, S.  C., Windhorst,  R. A.,  Phillips, S.,
Lanzetta, K., \& Yahil, A. 1998, ApJ, 496, L93
\bibitem[Eyles  \etal(2005)]{eyle05}  Eyles,  L.  P., Bunker,  A.  J.,
Stanway, E.  R., Lacy, M.,  Ellis R. S.,  \& Doherty, M.  2005, MNRAS,
364, 443
\bibitem[Eyles \etal(2007)]{eyle07} Eyles, L. P., Bunker, A. J., Ellis
R. S., Lacy,  M., Stanway, E. R., Stark, D., \&  Chiu, K. 2007, MNRAS,
374, 910
\bibitem[Fan  \etal(2006)]{fan06}  Fan, X.,  Strauss,  M. A.,  Becker,
R. H., et al.  2006, AJ, 132, 117
\bibitem[Georgakakis  \etal(2003)]{geor03}  Georgakakis, A.,  Hopkins,
A. M., Sullivan, M., Afonso, J., Georgantopoulos, I., Mobasher, B., \&
Cram, L. E. 2003, MNRAS, 345, 939
\bibitem[Giavalisco  \etal(2004)]{giav04}  Giavalisco, M.,  Dickinson,
M., Ferguson, H. C., et al. 2004, ApJ, 600, L103
\bibitem[Gonzaga \etal(2005)]{gonz05} Gonzaga,  S., et al. 2005, ``ACS
Instrument Handbook", Version 6.0, (Baltimore:STScI)
\bibitem[Guzman  \etal(1997)]{guzm97} Guzman,  R.,  Gallego, J.,  Koo,
D. C., Phillips,  A. C., Lowenthal, J. D.,  Faber, S. M., Illingworth,
G. D., \& Vogt, N. P. 1997 ApJ, 489, 559
\bibitem[Hathi  \etal(2007)]{hath07} Hathi,  N. P.,  Malhotra,  S., \&
Rhoads, J. E.  2007, ApJ, submitted (astro-ph/0709.0520)
\bibitem[Heavens   \etal(2004)]{heav04}  Heavens,   A.,   Panter,  B.,
Jimenez, R., \& Dunlop, J. 2004, Nature, 428, 625
\bibitem[Hu  \etal(2002)]{hu02}  Hu, E.  M.,  Cowie,  L. L.,  McMahon,
R. G., Capak, P., Iwamuro, F., Kneib, J.-P., Maihara, T., \& Motohara,
K. 2002, ApJ, 568, L75
\bibitem[Huchra \etal(1983)]{huch83} Huchra, J. P., Davis, M., Latham,
D., \& Tonry, J. 1983, ApJS, 52, 89
\bibitem[Hunter   \&  Elmegreen(2006)]{hunt06}   Hunter,  D.   A.,  \&
Elmegreen, B. G. 2006, ApJS, 162, 49
\bibitem[Jansen  \etal(2000a)]{jans00a}  Jansen,  R.  A.,  Franx,  M.,
Fabricant, D., \& Caldwell, N. 2000a, ApJS, 126, 271
\bibitem[Jansen \etal(2000b)]{jans00b}  Jansen, R. A.,  Fabricant, D.,
Franx, M., \& Caldwell, N. 2000b, ApJS, 126, 331
\bibitem[Jansen   \&  Kannappan(2001)]{jans01}   Jansen,   R.  A.   \&
Kannappan, S. J. 2001, Ap\&SS, 276, 1151
\bibitem[Kodaira  \etal(2003)]{kodi03}  Kodaira,  K.,  Taniguchi,  Y.,
Kashikawa, N., et al. 2003, PASJ, 55, L17
\bibitem[Kodama \etal(2004)]{koda04} Kodama,  T., Yamada, T., Akiyama,
M., et al. 2004, MNRAS, 350, 1005
\bibitem[Koekemoer  \etal(2002)]{koek02} Koekemoer,  A.  M., Fruchter,
A. S., Hook, R. N., \&  Hack, W. 2002, The 2002 \emph{HST} Calibration
Workshop,   ed.   S.   Arribas,   A.  Koekemoer,   and   B.   Whitmore
(Baltimore:STScI), 337
\bibitem[Kormendy(1977)]{korm77} Kormendy, J. 1977, ApJ, 218, 333
\bibitem[Kurk \etal(2004)]{kurk04} Kurk, J. D., Cimatti, A., di Serego
A, S., Vernet,  J., Daddi, E., Ferrara, A., \&  Ciardi, B. 2004, A\&A,
422, L13
\bibitem[Labb\'{e}  \etal(2006)]{labb06} Labb\'{e},  I.,  Bouwens, R.,
Illingworth, G. D., \& Franx, M. 2006, ApJ, 649, L67
\bibitem[Lynden-Bell(1967)]{lynd67} Lynden-Bell,  D. 1967, MNRAS, 136,
101
\bibitem[Maraston(2005)]{mara05} Maraston, C. 2005, MNRAS, 362, 799
\bibitem[Malhotra  \etal(2005)]{malh05} Malhotra,  S., Rhoads,  J. E.,
Pirzkal, N., et al. 2005, ApJ, 626, 666
\bibitem[Marzke \etal(1994)]{marz94}  Marzke, R. O., Huchra,  J. P. \&
Geller, M. J. 1994, ApJ, 428, 43
\bibitem[McCarthy \etal(2004)]{mcca04} McCarthy, P. J. 2004, BAAS, 36,
1555
\bibitem[Nandra  \etal(2002)]{nand02} Nandra,  K.,  Mushotzky, R.  F.,
Arnaud,  K.,  Steidel, C.  C.,  Adelberger,  K.  L., Gardner,  J.  P.,
Teplitz, H. I., \& Windhorst, R. A. 2002, ApJ, 576, 625
\bibitem[Oke \&  Gunn(1983)]{oke83} Oke, J.  B., \& Gunn, J.  E. 1983,
ApJ, 266, 713
\bibitem[Panter \etal(2007)]{pant07} Panter, B., Jimenez, R., Heavens,
A. F., \& Charlot, S. 2007, MNRAS, 378, 1550
\bibitem[Pascarelle \etal(1996)]{pasc96} Pascarelle, S. M., Windhorst,
R. A., Keel, W. C., \& Odewahn, S. C. 1996, Nature, 383, 45
\bibitem[Ravindranath     \etal(2006)]{ravi06}    Ravindranath,    S.,
Giavalisco,  M.,  Ferguson,  H.  C.,   et  al.  2006,  ApJ, 652, 963
\bibitem[Rhoads \etal(2004)]{rhoa04}  Rhoads, J.  E.,  Xu, C., Dawson,
S., et al. 2004, ApJ, 611, 59
\bibitem[Spergel  \etal(2007)]{sper07}  Spergel,   D.  N.,  Bean,  R.,
Dor\'{e}, O., et al. 2007, ApJS, 170, 377
\bibitem[Steidel  \etal(1999)]{stei99}  Steidel,  C.  C.,  Adelberger,
K. L., Giavalisco, M., Dickinson, M., \& Pettini, M. 1999, ApJ, 519, 1
\bibitem[Stern  \etal(2005)]{ster05} Stern, D.,  Yost, S.  A., Eckart,
M. E., Harrison,  F. A., Helfand, D. J.,  Djorgovski, S. G., Malhotra,
S., \& Rhoads, J. E. 2005, ApJ, 619, 12
\bibitem[Stetson(1987)]{stet87} Stetson, P. B. 1987, PASP, 99, 191
\bibitem[Taniguchi  \etal(2005)]{tani05}  Taniguchi,  Y.,  Ajiki,  M.,
Nagao, T., et al. 2005, PASJ, 57, 165
\bibitem[Taylor \etal(2007)]{tayl07} Taylor, V.  A., Conselice, C. J.,
Windhorst, R. A., \& Jansen, R. A. 2007, ApJ, 659, 162
\bibitem[van Albada(1982)]{vana82} van Albada, T. S. 1982, MNRAS, 201,
939
\bibitem[van   Dokkum  \etal(2003)]{vand03}   van   Dokkum,  P.    G.,
F\"{o}rster, S., Natascha, M., et al. 2003, ApJ, 587, L83
\bibitem[van  Dokkum \etal(2004)]{vand04} van  Dokkum, P.   G., Franx,
M., F\"{o}rster, S., et al. 2004, ApJ, 611, 703
\bibitem[Vanzella  \etal(2006)]{vanz06} Vanzella,  E.,  Cristiani, S.,
Dickinson, M., et al. 2006, A\&A, 454, 423
\bibitem[White  \etal(2007)]{whit07} White, R.   L., Helfand,  D.  J.,
Becker,  R. H.,  Glikman, E.,  \& de  Vries, W.   2007, ApJ,  654, 99
\bibitem[Williams \etal(1996)]{will96}  Williams, R. E.,  Blacker, B.,
Dickinson, M., et al. 1996, AJ, 112, 1335
\bibitem[Windhorst  \etal(1994)]{wind94}  Windhorst,  R.  A.,  Gordon,
J.  M., Pascarelle,  S. M.,  Schmidtke, P.  C., Keel,  W.  C., Burkey,
J. M., \& Dunlop, J. S. 1994, ApJ, 435, 577
\bibitem[Windhorst \etal(1998)]{wind98} Windhorst, R. A., Keel, W. C.,
\& Pascarelle, S. M. 1998, ApJ, 494, 27
\bibitem[Windhorst  \etal(2002)]{wind02}  Windhorst,  R.  A.,  Taylor,
V. A., Jansen, R. A., et al. 2002, ApJS, 143, 113
\bibitem[Windhorst \etal(2006)]{wind06} Windhorst, R. A., Cohen, S. H., 
Jansen, R. A., Conselice, C., \& Yan, H. 2006, NewAR, 50, 113
\bibitem[Windhorst  \etal(2007)]{wind07}   Windhorst,  R.  A.,  Hathi,
N. P.,  Cohen, S. H., \&  Jansen, R. A. 2007,  \emph{Advances in Space
Research}, in press (astro-ph/0703171)
\bibitem[Yan  \&  Windhorst(2004a)]{yan04a}   Yan,  H.  \&  Windhorst,
R. 2004a, ApJ, 600, L1
\bibitem[Yan  \&  Windhorst(2004b)]{yan04b}   Yan,  H.  \&  Windhorst,
R. 2004b, ApJ, 612, L93
\bibitem[Yan \etal(2005)]{yan05} Yan, H., Dickinson, M., Stern, D., et
al. 2005, ApJ, 634, 109
\bibitem[York \etal(2000)]{york00} York, D. G., et al. 2000, 120, 1579
\bibitem[Zibetti \etal(2004)]{zibe04} Zibetti, S., White, S. D. M., \&
Brinkmann, J.  2004, MNRAS, 347, 556
\end{thebibliography}
\end{document}